\documentclass[%
 reprint,
 amsmath,amssymb,
 aps,
]{revtex4-2}

\usepackage{graphicx}
\usepackage{dcolumn}
\usepackage{bm}
\usepackage[dvipsnames]{xcolor}

\begin{document}

\preprint{APS/123-QED}

\title{Influence of Many-Body Dipole-Dipole Interactions on Excitation Transfer in a Dense Gas}

\author{A. A. Bobrov}
\affiliation{Joint Institute for High Temperatures, Russian Academy of Sciences, Moscow 125412, Russia}
\email[]{abobrov@inbox.ru}
\author{S. A. Saakyan}
\affiliation{Joint Institute for High Temperatures, Russian Academy of Sciences, Moscow 125412, Russia}
\author{B. B. Zelener}
\affiliation{Joint Institute for High Temperatures, Russian Academy of Sciences, Moscow 125412, Russia}
\author{V. A. Sautenkov}
\affiliation{Joint Institute for High Temperatures, Russian Academy of Sciences, Moscow 125412, Russia}

\date{\today}

\begin{abstract}
We study non-radiative dipole-dipole induced excitation transfer in dense Rb vapour. We show that density dependence of characteristic time of the excitation diffusion changes from linear to non-linear for high Rb density. It is attributed to the breakdown of binary collisions approach to the dipole-dipole excitation transfer. It is shown that the observed result can be qualitatively described if the mean free path is replaced by mean interatomic distance for the characteristic diffusion length. 
\end{abstract}

\maketitle


Resonant energy transfer between quantum emitters, fundamentally mediated by dipole-dipole (DD) interactions, underpins crucial phenomena across physics~\cite{lewis1980}, chemistry~\cite{schafer2019modification}, and biology~\cite{piston2007fluorescent}, from photosynthesis~\cite{gtze2023excitation} to quantum information processing~\cite{lukin2001dipole}. The manifestation of these interactions in dense gases was first recognized in the context of spectral line shapes, where the self-broadening observed in early studies~\cite{Vlasov} was attributed to the perturbing influence of DD couplings. While the DD induced self-broadening in dense gases is actively studied (see e. g.~\cite{Lesanovsky, Christaller, Alaeian}), the role of DD coupling in facilitating nonradiative excitation transfer in dense neutral gases remains comparatively unexplored. This transfer mechanism is a nonradiative process wherein an excited atom (donor) de-excites by resonantly coupling to a nearby ground-state atom (acceptor), leading to an effective "hop" of the electronic excitation.

Of great interest in the recent time is the study of the DD interactions in dense gases when the gas number density $N$ becomes much larger than cube of the wavenumber $k^3$. The increase in density can drive a transition into a regime when the DD interactions cannot be reduced to binary collisions and the many-body interactions and collective effects should be taken into account. Recent studies~\cite{Lesanovsky, sautenkov2025optical,liang} show that there are strong indications that the many-body DD interactions significantly affect the spectral line shifts and shapes in a dense gas.

In this Letter we focus on study of density effects in the DD mediated excitation transport. We were able to experimentally observe the breakdown of the binary collisions regime of the DD mediated excitation transfer which was accompanied by the sharp change in the density dependence of the excitation transport dynamics from a linear to a strongly non-linear.

In a binary collisions approximation the DD excitation transport is usually characterized by the mean free path $l_d$ which an excited atom goes through in a gas before the excitation "hops" to a ground-state atom. The mean free path can be expressed using cross-section of the resonance excitation exchange $\sigma_{re}$ as $l_d=v_{th} (\langle\sigma_{re} v\rangle N)^{-1}$, where $v_{th}$ is the most probable speed of the atom in the gas and the averaging is made over thermal distribution of  the atomic velocities $v$. The excitation-exchange cross-section is closely connected to the DD "collision" cross-section $\sigma$, which determines the self-broadening $\Gamma$ in a gas: $\Gamma=\langle\sigma v\rangle N$. It was shown, both theoretically and experimentally~\cite{van1997dipole}, that in the binary collision approximation roughly every second DD "collision" leads to the excitation exchange, so the mean free path is simply related to the self-broadening $\Gamma$ as $l_d\approx2v_{th}/\Gamma$.

Using the mean free path expression, the diffusion coefficient for the DD excitation transfer then can be introduced~\cite{Zajonc1981, van1997dipole}:
\begin{eqnarray}
D=\frac{l_d^2}{2(\langle\sigma_{re}\rangle N)^{-1}}\approx\frac{v_{th}^2}{\Gamma}.
\end{eqnarray}

While the equation (1) was successfully used to describe self-broadening of selective reflection spectrum for gas densities where the DD interactions dominate~\cite{van1997dipole}, it, however, gives incorrect result of zero diffusion in the limit of infinite density. The breakdown occurs due to the linear dependence of the probability of the excitation exchange on density. For large enough densities the probability becomes so high, that the excitation "hops" from the excited atom to the ground state one much faster than the excited atom can propagate a fraction of the interatomic distance, and the above defined mean free path becomes meaningless.

\begin{figure*}[t]
  \includegraphics[width=1\textwidth]{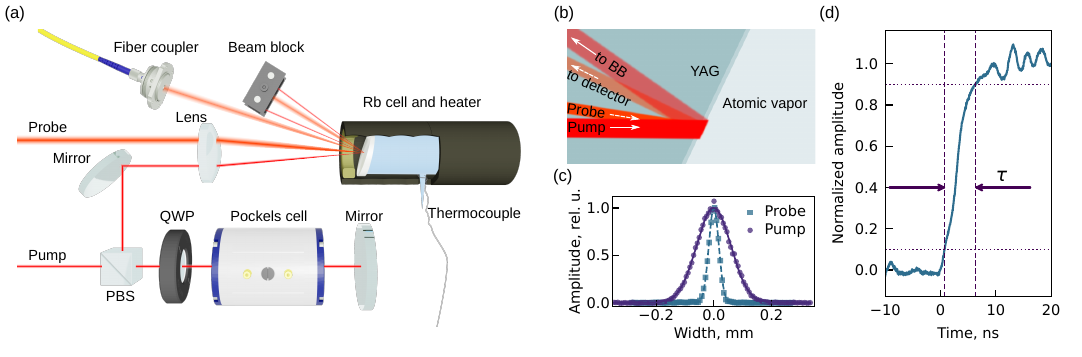}
  \caption{\label{fig:1} (a)~ Experimental setup. (b)~Incidence scheme of the probe, pump and reflected beams. (c)~The pump (blue circles) and probe (green squares) horizontal profile. The solid green and dashed blue curves are the best fit using a Gaussian function with $w_{pump}^H=119$~$\mu$m  and $w_{probe}^H=35$~$\mu$m. 
(d)~Typical rise of the reflected probe signal after the pump is switched
off at $N=1.4\times10^{16}$~cm$^{-3}$.
The measured 10--90\% rise time is $\tau=5.5$~ns.}
\end{figure*}

For the correct description of the DD excitation transfer for higher densities the binary collision approximation is not enough and the many-body interactions should be taken into account. Rigorous quantitative study of the many-body problem is rather complicated and require extensive calculations~\cite{leegwater1994self}. However, qualitatively, we can simplify the description of the DD excitation transfer for high gas densities if we assume that the transfer is still diffusion-like at least for some region of densities~\cite{Phelps1970}. Then it is natural to assume, for dimensional reasons, that the mean interatomic distance $r_a=(4\pi N/3)^{-1/3}$ should be used as mean free path for the excitation transfer, i. e. $l_d\sim r_a$, when $v_{th}/\Gamma\lesssim r_a$. If we assume that characteristic diffusion time remains of the order of $\Gamma^{-1}$, the diffusion coefficient should became proportional to $N^{1/3}$:
\begin{eqnarray}
D\sim r_a^2\Gamma\sim N^\frac{1}{3}.
\end{eqnarray}
To our knowledge, except for the brief notice by Phelps and Chen~\cite{Phelps1970}, there are no experimental or theoretical investigations of this sharp change of density dependence of the non-radiative excitation diffusion coefficient from $\sim 1/N$, which follows from eq. (1), to $\sim N^{1/3}$, and here we present strong experimental evidence, that this transition indeed take place in a dense gas. 

In our experiment, we study excitation transfer in dense Rb vapor using the
setup shown in Fig.~\ref{fig:1}(a).
Measurements are performed in a high-temperature sapphire cell with YAG
windows. 
The Rb vapor density is varied by changing the cell temperature and is
determined from the measured temperature of the coldest spot of the cell,
which ranges from 280 to 384~$^\circ$C, corresponding to
$1.4\times10^{16}$--$1.6\times10^{17}$~cm$^{-3}$~\cite{alcock1984vapour}.

The excitation is created by a strong pump laser beam driving the $5S_{1/2}-5P_{3/2}$ transition. To monitor the atomic excitation a weak probe laser beam is used, which is reflected from the YAG--atomic-vapor interface.
The pump and probe beams are focused by a 350~mm focal length lens onto the
inner surface of the YAG window, as shown in Fig.~\ref{fig:1}(b).
The pump (probe) horizontal (H) and vertical (V) beam radii at the cell
interface are
$w_{\mathrm{pump}}^{\mathrm{H,V}}=\{119,97\}~\mu\mathrm{m}$
($w_{\mathrm{probe}}^{\mathrm{H,V}}=\{35,40\}~\mu\mathrm{m}$),
defined at the $1/e^2$ intensity level.
The pump intensity is 191~W/cm$^2$, the probe intensity is maintained well below the pump intensity. The measured horizontal beam profiles are shown in Fig.~\ref{fig:1}(c).

The pump and probe beams have the same linear polarization.
The probe laser frequency is blue-detuned by 3~GHz from the $^{85}$Rb
$5S_{1/2}(F=3)-5P_{3/2}(F'=4)$ transition, while the pump laser
is red-detuned by 4~GHz from this transition and monitored with a wavemeter.
This detuning suppresses coherent four-wave mixing near zero pump--probe
detuning~\cite{sautenkov2023coherent}.

To study the excitation transfer dynamics, the pump radiation is rapidly switched
off using a Pockels cell in a retro-reflected configuration.
The Pockels cell is operated at approximately quarter-wave voltage and has
a measured 10--90\% switching time of less than 2~ns.

The reflected probe beam is coupled into a 2 meter long optical fiber for
spatial filtering and reduction of electrical pickup from the
Pockels-cell driver.
The fiber output is detected by an avalanche photodiode with a
0.5~GHz bandwidth, and the signal is recorded with an oscilloscope
having a 0.5~GHz bandwidth.
The Pockels-cell switching time is measured using the same detection
system.
The contribution of the finite detection bandwidth to the measured
rise times is estimated to be below 0.5~ns using the root-sum-square
approximation.

\begin{figure}[t]
  \includegraphics[width=1\columnwidth]{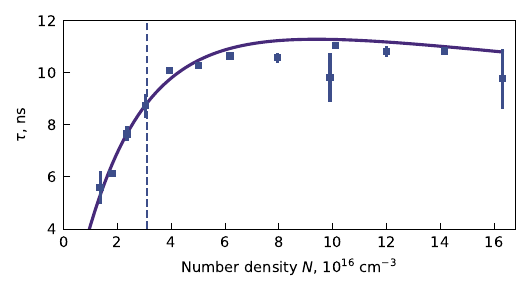}
  \caption{\label{fig:2} The probe reflection signal rise time after the pump switch off versus Rb vapour density. 
  Points are the experimental results, solid line is the results of fit by the eq.~(5), 
  dashed line is the critical density $N_{cr}=3.1\times10^{16}$~cm$^{-3}$, at which the 
  distance an atom travels during time $\sim\Gamma^{-1}$ is of order of the average interatomic
  distance.}
\end{figure}

After the pump is switched off, the reflected probe intensity increases.
A typical transient at $N=1.4\times10^{16}$~cm$^{-3}$ is shown in
Fig.~\ref{fig:1}(d).
We characterize the transient by its 10--90\% rise time $\tau$, defined as the
interval between 10\% and 90\% of the total signal change.
The main experimental result is the dependence of this rise time on the
Rb vapor density, shown in Fig.~\ref{fig:2}.

The observed dependence of the probe rise time can be described in the following way. The pump radiation decreases the ground state population of the Rb atoms near the window surface, that in turn decreases the probe beam reflection signal. The probe reflection signal is formed in a thin layer of thickness $l_a\sim\lambda$~\cite{Boyd}. When the pump radiation is turned off, the excited states diffuse via DD hopping towards the window and are quenched in collisions with the window’s inner surface. This decay of the excited states population in turn leads to the sharp rise of the probe reflected signal. The rise time of the signal is determined by the diffusion time $\tau_d$, which can be qualitatively estimated using the length $l_a$ and the diffusion coefficient $D$ as
\begin{eqnarray}
\tau_d\sim \frac{l_a^2}{D}.
\end{eqnarray}

For low density, where the equation (1) is valid, the rise time should have linear dependence on density $\tau_d\sim N$, which indeed can be seen in Figure~\ref{fig:2}. Then, with the increase in density, the slope steepness changes dramatically. This corresponds to change in the underlying diffusion: the excitation mean free path  becomes independent on the atoms motion. The slope for high densities follows much weaker dependence of the diffusion coefficient on density (2), which yields the density dependence of the rise  time $\tau_d\sim N^{-1/3}$.

At the intermediate density we can assume that both diffusion processes exist simultaneously. We can qualitatively describe the resulting diffusion coefficient as a sum of the motion-dominated coefficient and the coefficient determined only by hops of excitation. The square of the effective mean free path then will be equal to the sum of the squares of the motion mean free path and the mean interatomic distance:
\begin{eqnarray}
{l_d^*}^2\sim\left(\frac{2v_{th}}{\Gamma}\right)^2+const\cdot r_a^2.
\end{eqnarray}

We can further construct a fit formula for the diffusion time:
\begin{eqnarray}
\tau^*=C_1 (\lambda/2\pi)^2\left[\left(\frac{2v_{th}}{\Gamma}\right)^2+C_2 r_a^2\right]^{-1}\frac{4}{\Gamma},
\end{eqnarray}
where $C_1$ and $C_2$ are dimensionless fit parameters. We see from Figure~\ref{fig:2} that fitting of the measured rise time density dependence by equation (5) gives good description of the experimental data. To estimate the $\Gamma$ we used the expression $\Gamma=KN$ with the factor $K/2\pi\approx1.1\times10^{-16}$~GHz cm$^3$~\cite{lewis1980}. The resulting least squares fit coefficients values are $C_1=8.85$ and $C_2=3.30$. From the fit we can estimate the critical density value $N_{cr}\approx3.1\times10^{16}$~cm$^{-3}$, where mean free path $2v_{th}/\Gamma$ is equal to $\sqrt{C_2}r_a$ (see vertical line in Fig.~\ref{fig:2}). This critical density value qualitatively define the boundary beyond which the binary collisions approximation became invalid.

In conclusion, we have experimentally confirmed the transition of the density dependence of dipolar excitation diffusion from a linear to a significantly nonlinear regime, which is attributed to breakdown of binary collisions approach. We propose simple model, in which we substitute mean interatomic distance as mean free path, assuming the excitation transfer is diffusion-like for high density gas, and it yields satisfactory qualitative agreement with the experiment. For quantitative agreement, probably, it is necessary to account for many body interactions and to rigorously account for inhomogeneous excitation in the medium. This is beyond the scope of this letter and is in our plans for further research.

Another fascinating feature of our results is the very rapid quenching of the excitation. The observed times of 5 to 10 ns are much  lower than the spontaneous lifetime, which for the respected transition is 26 ns. This short excitation lifetime is consistent with our previous findings on spectral widths~\cite{bobrov2021dipole}, although accurate account of its contribution to the spectral width of the selective reflection spectrum requires separate study.

The short life time can pave the way for rapid control of the DD interactions in dense gases. The switching time of several nanoseconds is comparable to the recent study~\cite{Christaller}, where the DD interactions were controlled by rapid increase of the Rb vapour density in a nanocell by laser desorption of Rb atoms from the cell window inner surface.

Varying of the mean free path dependence on density may also be important for manipulating dipole-dipole interactions in confined geometries, which is also being actively studied~\cite{Alaeian,sargsyan2023}.

Our results show that dense gases can serve as a platform for studying collective quantum optical phenomena and non-equilibrium energy transport in a thermally active, disordered medium.

\begin{acknowledgments}
This work was supported by the Ministry of Science and Higher Education of the Russian Federation (State Assignment No. 075-00270-26-00). A.B. would like to express his gratitude to Julia Bobrova for her help with data processing.
\end{acknowledgments}


\bibliography{BiblioZotero3}

@article{lewis1980,
  title = {Collisional relaxation of atomic excited states, line broadening and
           interatomic interactions},
  author = {Lewis, EL},
  journal = {Physics Reports},
  volume = {58},
  number = {1},
  pages = {1--71},
  year = {1980},
  publisher = {Elsevier},
}

@article{schafer2019modification,
  title = {Modification of excitation and charge transfer in cavity
           quantum-electrodynamical chemistry},
  author = {Sch{\"a}fer, Christian and Ruggenthaler, Michael and Appel, Heiko
            and Rubio, Angel},
  journal = {Proceedings of the National Academy of Sciences},
  volume = {116},
  number = {11},
  pages = {4883--4892},
  year = {2019},
  publisher = {National Academy of Sciences},
}

@article{gtze2023excitation,
  title = {Excitation energy transfer between higher excited states of
           photosynthetic pigments: 2. Chlorophyll b is a B band excitation trap},
  author = {G{\"o}tze, Jan P and Lokstein, Heiko},
  journal = {Acs Omega},
  volume = {8},
  number = {43},
  pages = {40015--40023},
  year = {2023},
  publisher = {ACS Publications},
}

@article{piston2007fluorescent,
  title = {Fluorescent protein FRET: the good, the bad and the ugly},
  author = {Piston, David W and Kremers, Gert-Jan},
  journal = {Trends in biochemical sciences},
  volume = {32},
  number = {9},
  pages = {407--414},
  year = {2007},
  publisher = {Elsevier},
}

@article{lukin2001dipole,
  title = {Dipole blockade and quantum information processing in mesoscopic
           atomic ensembles},
  author = {Lukin, Mikhail D and Fleischhauer, Michael and Cote, Robin and Duan,
            LuMing and Jaksch, Dieter and Cirac, J Ignacio and Zoller, Peter},
  journal = {Physical review letters},
  volume = {87},
  number = {3},
  pages = {037901},
  year = {2001},
  publisher = {APS},
}

@article{Vlasov,
  author = {Fursov, V S and Vlasov, A A},
  journal = {J. Exp. Theor. Phys.},
  volume = {6},
  pages = {750},
  year = {1936},
}

@article{Lesanovsky,
  author = {Lesanovsky, I and Olmos, B. and Guerin, W and Kaiser, R},
  journal = {Phys. Rev. {\rm A}},
  volume = {100},
  pages = {021401},
  year = {2019},
}

@article{Christaller,
  author = {Christaller, F and others},
  journal = {Phys. Rev. Lett.},
  volume = {128},
  pages = {173401},
  year = {2022},
}

@article{Alaeian,
  author = {Alaeian, H and Skljarow, A and Scheel, S and Pfau, T and Löw, R},
  journal = { New J. Phys.},
  volume = {26},
  pages = {055001},
  year = {2024},
}

@article{van1997dipole,
  author = {{Van Kampen}, H and Sautenkov, V A and Shalagin, A M and Eliel, E R
            and Woerdman, J P},
  journal = {Phys. Rev. {\rm A}},
  volume = {56},
  number = {5},
  pages = {3569},
  year = {1997},
}

@article{Zajonc1981,
  author = {Zajonc, A G and Phelps, A V },
  journal = {Phys. Rev. {\rm A}},
  volume = {23},
  number = {5},
  pages = {2479},
  year = {1981},
}

@article{Phelps1970,
  author = {Phelps, A V and Chen, C L },
  journal = {Bull. Am. Phys. Soc. },
  volume = {15},
  pages = {428},
  year = {1970},
}

@article{sautenkov2025optical,
  title = {Optical-field-induced dips and splits in nonlinear spectra of
           selective reflection from high-density atomic vapor},
  author = {Sautenkov, Vladimir and Saakyan, Sergey and Bobrov, Andrei and
            Zelener, Boris B},
  journal = {Journal of Quantitative Spectroscopy and Radiative Transfer},
  pages = {109796},
  year = {2025},
  publisher = {Elsevier},
}

@article{leegwater1994self,
  title = {Self-broadening and exciton line shifts in gases: Beyond the
           local-field approximation},
  author = {Leegwater, Jan A and Mukamel, Shaul},
  journal = {Physical Review A},
  volume = {49},
  number = {1},
  pages = {146},
  year = {1994},
  publisher = {APS},
}

@article{bobrov2021dipole,
  title = {Dipole--dipole broadening in the selective reflection of an intense
           laser beam from the interface between a transparent dielectric and a
           dense resonance gas},
  author = {Bobrov, Andrei Aleksandrovich and Saakyan, Sergei Aramovich and
            Sautenkov, Vladimir Alekseevich and Zelener, Boris Borisovich},
  journal = {JETP Letters},
  volume = {114},
  number = {9},
  pages = {524--527},
  year = {2021},
  publisher = {Springer},
}

@book{Boyd,
  author = "Boyd, Robert W",
  title = "{Nonlinear optics}",
  publisher = "Academic Press",
  address = "Amsterdam",
  year = "2003",
  url = "https://cds.cern.ch/record/640132",
}

@article{liang,
  title = {Optical two-dimensional coherent spectroscopy of many-body
           dipole--dipole interactions and correlations in atomic vapors},
  author = {Liang, Danfu and Li, Hebin},
  journal = {The Journal of chemical physics},
  volume = {154},
  number = {21},
  year = {2021},
  publisher = {AIP Publishing},
}

@article{sargsyan2023,
  title = {Competing van der Waals and dipole-dipole interactions in optical
           nanocells at thicknesses below 100 nm},
  author = {Sargsyan, Armen and Momier, Rodolphe and Leroy, Claude and Sarkisyan
            , David},
  journal = {Physics Letters A},
  volume = {483},
  pages = {129069},
  year = {2023},
  publisher = {Elsevier},
}

@article{sautenkov2023coherent,
  title = {Coherent resonances in a dipole-broadened contour of selective
           reflection from the transparent insulator--atomic rubidium vapor
           interface},
  author = {Sautenkov, VA and Saakyan, SA and Bobrov, AA and Vilshanskaya, EV
            and Zelener, BB},
  journal = {Bulletin of the Lebedev Physics Institute},
  volume = {50},
  number = {Suppl 5},
  pages = {S599--S605},
  year = {2023},
  publisher = {Springer},
}

@article{alcock1984vapour,
  title = {Vapour pressure equations for the metallic elements: 298--2500K},
  author = {Alcock, CB and Itkin, VP and Horrigan, MK},
  journal = {Can. Metall. Q.},
  volume = {23},
  number = {3},
  pages = {309--313},
  year = {1984},
  publisher = {Taylor \& Francis},
  doi = {10.1179/cmq.1984.23.3.309},
}

\end{document}